\documentstyle[preprint,epsfig,aps,amsmath,floats,tighten]{revtex}

\newcommand{\etal}{{\it et al.}}
\newcommand{\mt}{m_T}
\newcommand{\ppbar}{p\overline{p}}
\newcommand{\Dzero}{D\O}
\newcommand{\dzero}{\Dzero}
\newcommand{\mw}{M_{W}}
\newcommand{\mz}{M_{Z}}
\newcommand{\qqbar}{q\overline{q}}
\newcommand{\pt}{p_T}
\newcommand{\mpt}{\mbox{$\rlap{\kern0.1em/}\pt$}}
\newcommand{\pe}{p(e)}
\newcommand{\pev}{\vec\pe}
\newcommand{\pte}{\pt(e)}
\newcommand{\ptev}{\vec\pte}
\newcommand{\utv}{\vec\ut}
\newcommand{\ut}{u_T}
\newcommand{\ptnu}{\pt(\nu)}
\newcommand{\ptnuv}{\vec\ptnu}
\newcommand{\ptw}{\pt(W)}
\newcommand{\ptwv}{\vec\ptw}
\newcommand{\wev}{W\to e\nu}
\newcommand{\zee}{Z\to ee}
\newcommand{\wte}{W\to \tau\nu\to e\nu\overline\nu\nu}
\newcommand{\cem}{c_{\rm EM}}

\newcommand{\gt}{>}
\newcommand{\alphaem}{\alpha_{\rm EM}}
\newcommand{\deltaem}{\delta_{\rm EM}}
\newcommand{\alphamb}{\alpha_{\rm mb}}
\newcommand{\srec}{s_{\rm rec}}
\newcommand{\ptee}{\pt(ee)}
\newcommand{\pteev}{\vec\ptee}

\newcommand{\ueta}{u_\eta}

\newcommand{\pteta}{p_\eta (ee)}
\newcommand{\GEAN}{{\sc geant}}
\newcommand{\alpharec}{\alpha_{\rm rec}}
\newcommand{\betarec}{\beta_{\rm rec}}
\newcommand{\PM}{\pm}

\newcommand{\phie}{\phi(e)}
\newcommand{\PRL}{Phys. Rev. Lett.}
\newcommand{\PL}{Phys. Lett.}
\newcommand{\PR}{Phys. Rev.}
\newcommand{\NP}{Nucl. Phys.}
\newcommand{\NIM}{Nucl. Instrum. Methods in Phys. Res.}
\newcommand{\ZP}{Z.~Phys.}

\begin{document}
\lefthyphenmin=2
\righthyphenmin=3
\title{ \bf A New Measurement of the W Boson Mass at D\O}
%
\author{\centerline{The D\O\ Collaboration
  \thanks{Submitted to the {\it International Europhysics Conference} on
         {\it High Energy Physics}, {\it EPS-HEP99},
          \hfill\break
           15 -- 21 July, 1999, Tampere, Finland.}}}
\address{
\centerline{Fermi National Accelerator Laboratory, Batavia, Illinois 60510}
}
\date{\today}

\maketitle

\begin{abstract}
We present a new measurement of the $W$ mass using the
 $W\!\rightarrow \! e\nu$ data from the D\O\ forward detectors at the
 Fermilab Tevatron $p \bar p$ Collider. This is the first
 measurement of the $W$ mass with electron candidates in the range
 $1.5 < \mid \eta \mid < 2.5$.  We present measurements of the $W$
 mass using the transverse mass, the electron transverse momentum and
 the neutrino transverse momentum, and the combined result using all
 three techniques. The combination of the forward detector measurement
 with the previous measurements using the central detector gives a new
 precise measurement of the $W$ mass from D\O.
\end{abstract}
\newpage
\begin{center}
%
B.~Abbott,$^{45}$                                                             
M.~Abolins,$^{42}$                                                            
V.~Abramov,$^{18}$                                                            
B.S.~Acharya,$^{11}$                                                          
I.~Adam,$^{44}$                                                               
D.L.~Adams,$^{54}$                                                            
M.~Adams,$^{28}$                                                              
S.~Ahn,$^{27}$                                                                
V.~Akimov,$^{16}$                                                             
G.A.~Alves,$^{2}$
% ...
\\                                                                            
\vskip 0.30cm                                                                 
\centerline{(D\O\ Collaboration)}                                             
\vskip 0.30cm                                                                 
\centerline{$^{1}$Universidad de Buenos Aires, Buenos Aires, Argentina}       
\centerline{$^{2}$LAFEX, Centro Brasileiro de Pesquisas F{\'\i}sicas,         
                  Rio de Janeiro, Brazil}                                     
\centerline{$^{3}$Universidade do Estado do Rio de Janeiro,                   
                  Rio de Janeiro, Brazil}                                     
\centerline{$^{4}$Institute of High Energy Physics, Beijing,                  
                  People's Republic of China}                                 
\centerline{$^{5}$Universidad de los Andes, Bogot\'{a}, Colombia}             
\centerline{$^{6}$Universidad San Francisco de Quito, Quito, Ecuador}         
\centerline{$^{7}$Institut des Sciences Nucl\'eaires, IN2P3-CNRS,             
                  Universite de Grenoble 1, Grenoble, France}                 
\centerline{$^{8}$DAPNIA/Service de Physique des Particules, CEA, Saclay,     
                  France}                                                     
\centerline{$^{9}$Panjab University, Chandigarh, India}                       
\centerline{$^{10}$Delhi University, Delhi, India}                            
\centerline{$^{11}$Tata Institute of Fundamental Research, Mumbai, India}     
\centerline{$^{12}$Kyungsung University, Pusan, Korea}                        
\centerline{$^{13}$Seoul National University, Seoul, Korea}                   
\centerline{$^{14}$CINVESTAV, Mexico City, Mexico}                            
\centerline{$^{15}$Institute of Nuclear Physics, Krak\'ow, Poland}            
\centerline{$^{16}$Institute for Theoretical and Experimental Physics,        
                   Moscow, Russia}                                            
\centerline{$^{17}$Moscow State University, Moscow, Russia}                   
\centerline{$^{18}$Institute for High Energy Physics, Protvino, Russia}       
\centerline{$^{19}$Lancaster University, Lancaster, United Kingdom}           
\centerline{$^{20}$University of Arizona, Tucson, Arizona 85721}              
\centerline{$^{21}$Lawrence Berkeley National Laboratory and University of    
                   California, Berkeley, California 94720}                    
\centerline{$^{22}$University of California, Davis, California 95616}         
\centerline{$^{23}$University of California, Irvine, California 92697}        
\centerline{$^{24}$University of California, Riverside, California 92521}     
\centerline{$^{25}$Florida State University, Tallahassee, Florida 32306}      
\centerline{$^{26}$University of Hawaii, Honolulu, Hawaii 96822}              
\centerline{$^{27}$Fermi National Accelerator Laboratory, Batavia,            
                   Illinois 60510}                                            
\centerline{$^{28}$University of Illinois at Chicago, Chicago,                
                   Illinois 60607}                                            
\centerline{$^{29}$Northern Illinois University, DeKalb, Illinois 60115}      
\centerline{$^{30}$Northwestern University, Evanston, Illinois 60208}         
\centerline{$^{31}$Indiana University, Bloomington, Indiana 47405}            
\centerline{$^{32}$University of Notre Dame, Notre Dame, Indiana 46556}       
\centerline{$^{33}$Purdue University, West Lafayette, Indiana 47907}          
\centerline{$^{34}$Iowa State University, Ames, Iowa 50011}                   
\centerline{$^{35}$University of Kansas, Lawrence, Kansas 66045}              
\centerline{$^{36}$Kansas State University, Manhattan, Kansas 66506}          
\centerline{$^{37}$Louisiana Tech University, Ruston, Louisiana 71272}        
\centerline{$^{38}$University of Maryland, College Park, Maryland 20742}      
\centerline{$^{39}$Boston University, Boston, Massachusetts 02215}            
\centerline{$^{40}$Northeastern University, Boston, Massachusetts 02115}      
\centerline{$^{41}$University of Michigan, Ann Arbor, Michigan 48109}         
\centerline{$^{42}$Michigan State University, East Lansing, Michigan 48824}   
\centerline{$^{43}$University of Nebraska, Lincoln, Nebraska 68588}           
\centerline{$^{44}$Columbia University, New York, New York 10027}             
\centerline{$^{45}$New York University, New York, New York 10003}             
\centerline{$^{46}$University of Rochester, Rochester, New York 14627}        
\centerline{$^{47}$State University of New York, Stony Brook,                 
                   New York 11794}                                            
\centerline{$^{48}$Brookhaven National Laboratory, Upton, New York 11973}     
\centerline{$^{49}$Langston University, Langston, Oklahoma 73050}             
\centerline{$^{50}$University of Oklahoma, Norman, Oklahoma 73019}            
\centerline{$^{51}$Brown University, Providence, Rhode Island 02912}          
\centerline{$^{52}$University of Texas, Arlington, Texas 76019}               
\centerline{$^{53}$Texas A\&M University, College Station, Texas 77843}       
\centerline{$^{54}$Rice University, Houston, Texas 77005}                     

\end{center}

\vfill\eject

In the standard model of the electroweak interactions (SM)~\cite{sm}, the mass
of the $W$ boson is predicted to be 
\begin{equation}
\mw = \left( \frac{\pi\alpha(\mz^2)}
 {\sqrt{2}G_F}\right)^\frac{1}{2} \frac{1}{\sin\theta_w\sqrt{1-\Delta r_W}}\; .
\label{eq:dr}
\end{equation}
In the ``on-shell'' scheme~\cite{on_shell} $\cos\theta_w = \mw/\mz$, 
where $\mz$ is the $Z$ boson mass.  A measurement of $\mw$,
together with $\mz$, the Fermi constant ($G_F$), and the electromagnetic 
coupling constant ($\alpha$),
determines the electroweak radiative corrections $\Delta r_W$ experimentally. 
Purely electromagnetic corrections are absorbed into the value of $\alpha$
by evaluating it at $Q^2=\mz^2$~\cite{amz}. 
The dominant contributions to $\Delta r_W$
arise from loop diagrams that involve the top quark and the Higgs boson. 
If additional particles which couple to the $W$ boson exist, they give
rise to additional contributions to $\Delta r_W$. Therefore, a measurement of
$\mw$ is one of the most stringent experimental tests of SM predictions.
Deviations from the predictions may indicate the existence of new physics.
Within the SM, measurements of $\mw$ and the mass of the top quark constrain  
the mass of the Higgs boson.

This paper reports a new measurement of the $W$ boson mass using electrons 
detected in forward calorimeters. We used data  recorded by the \Dzero\ detector~\cite{d0_nim} during the
1994--1995 run of the Fermilab Tevatron $\ppbar$ collider. This forward 
electron measurement complements our previous
 measurement with central electrons \cite{wmass1bcc} because the more complete
 rapidity coverage gives useful constraints on model parameters that permit
 reduction of the systematic error, in addition to increasing the
  statistical
 precision. Combining this measurement with our previously published  measurements with central electrons using data taken in 1992--1993 and 1994--1995, determines the $W$ boson mass to a precision of 93~MeV.

At the Tevatron, $W$ bosons are produced mainly through $\qqbar$
annihilation. We detect them by their decays into electron-neutrino pairs,
characterized by an isolated electron~\cite{ep} with large transverse momentum
($\pt$) and significant transverse momentum imbalance ($\mpt$). 
The $\mpt$ is due to the neutrino which escapes detection. 
Many other particles of lower momenta, which
recoil against the $W$ boson, are produced in the
breakup of the proton and antiproton.
We refer to them collectively as the underlying event.

At the trigger level we require $\mpt>15$~GeV and an energy cluster in the
electromagnetic (EM) calorimeter with $\pt>20$~GeV.
The cluster must be isolated and have a shape consistent with that of an 
electron shower.

During event reconstruction, electrons are identified as energy clusters in the
EM calorimeter which satisfy isolation and shower shape cuts and
have a drift chamber track pointing towards the cluster centroid.
We determine forward electron  energies by adding the 
energy depositions in the first
$\approx40$ radiation lengths of the calorimeter in a cone of radius 20 cm, centered on
the  highest energy deposit in the cluster. Fiducial cuts reject electron
candidates near calorimeter module edges and ensure a uniform calorimeter
response for the selected electrons. The electron momentum ($\pev$) is 
determined by combining its energy with its direction which is
obtained from the shower centroid position and the drift chamber track.
The trajectories of the electron and the proton beam define the position of the
event vertex. 

We measure the sum of the transverse momenta
of all the particles recoiling against the $W$ boson, $\utv = \sum_i E_i
\sin\theta_i \hat u_i$, where $E_i$ is the energy deposition in the 
$i^{th}$ calorimeter cell and $\theta_i$ is the angle defined by the cell 
center, the event vertex, and the proton beam. The unit vector $\hat u_i$ points
perpendicularly from the beam to the cell center. 
The calculation of $\utv$ excludes the cells occupied by the electron.
The sum of the momentum components along the beam is not well measured because
of particles escaping through the beam pipe. 
From momentum conservation we infer the transverse neutrino momentum, 
$\ptnuv = -\ptev-\utv$, and the transverse momentum of the $W$ boson,
$\ptwv=-\utv$. 

We select a $W$ boson sample of 11{,}090 events by requiring 
$\ptnu>30$~GeV, $\ut<15$~GeV, and an electron candidate with 
$1.5<|\eta|<2.5$ and $\pte>30$~GeV.

Since we do not measure the longitudinal momentum components of the neutrinos
from $W$ boson decays, we cannot reconstruct the $e\nu$ invariant mass.
Instead, we extract the $W$ boson mass from the spectra of the electron $\pt$
and the transverse mass, $\mt = \sqrt{2\pte\ptnu(1-\cos\Delta\phi)}$,
where $\Delta\phi$ is the azimuthal separation between the two leptons.
We perform a maximum likelihood fit to the data using probability density
functions from a Monte Carlo program.
Since neither $\mt$ nor $\pte$ are Lorentz invariants, we have to model
the production dynamics of $W$ bosons to correctly predict the spectra.
The $\mt$ spectrum is insensitive to transverse boosts at leading order in
$\ptw/\mw$ and is therefore less sensitive to the $W$ boson production model
than the $\pte$ spectrum. On the other hand, the $\mt$ spectrum depends strongly
on the detector response to the underlying event and is therefore more sensitive
to detector effects than the $\pte$ spectrum. The shape of the neutrino $\ptnu$
 spectrum is sensitive to both the $W$ boson production dynamics and the recoil
momentum measurement. By performing the measurement using all three spectra we
 provide a powerful cross check with complementary systematics.

$Z$ bosons decaying to electrons provide an important control
sample. We use them to calibrate the
detector response to the underlying event and to the electrons, and to 
constrain the model for intermediate vector boson production used in the
Monte Carlo simulations.  

A $\zee$ event is characterized by two isolated high-$\pt$ electrons.
We trigger on events with at least two EM clusters with $\pt\gt 20$~GeV. 
We accept $Z\rightarrow ee$ decays with at least one forward electron in
 the pseudorapidity range $1.5<|\eta|<2.5$, where $\eta = - \ln \left( \tan \frac{\theta}{2} \right)$, and the other electron to be either forward or central with pseudorapidity $|\eta|<1.0$. The central electron is required 
to have $\pt>25$~GeV but is allowed not to have a matching drift chamber 
track. The forward electron candidate is required to have  $\pt>30$~GeV and a matching drift chamber track. This selection accepts 1{,}687 events.

For this measurement we used a fast Monte Carlo program developed  for our  central electron analysis  \cite{wmass1bcc}, with some modifications in the simulation of forward electron events. The program 
generates $W$ and $Z$ bosons with the
rapidity and $\pt$ spectra given by a calculation using soft gluon 
resummation~\cite{ly} and the MRSA~\cite{mrsa} parton distribution 
functions. 
The line shape is a relativistic Breit-Wigner, skewed by the mass
dependence of the parton luminosity. The measured intrinsic 
widths~\cite{width,ewwg} are used.
The angular distribution of the decay electrons includes a $\ptw$-dependent 
${\cal O}(\alpha_s)$ correction~\cite{mirkes}. The program also generates
$\wev\gamma$~\cite{berends}, $\zee\gamma$~\cite{berends}, and $\wte$ decays.

The program smears the generated $\pev$ and $\utv$ vectors using a
parameterized  detector response model and applies inefficiencies introduced by
the trigger  and event selection requirements. 
The model parameters are adjusted to match the data and are discussed below.

\begin{figure}[htb]
\begin{tabular}{c}
\includegraphics{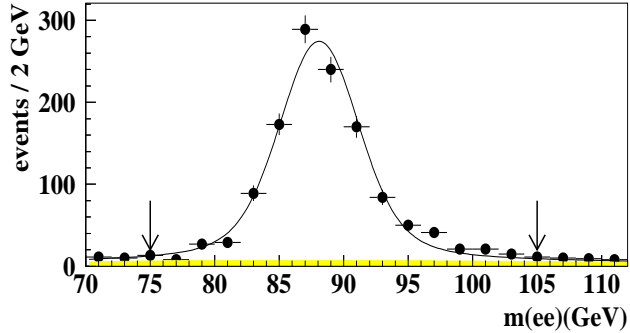}
\end{tabular}
\vskip -0.25 cm
%
%
\caption[]{The dielectron invariant mass distribution of the $Z$ data  ($\bullet$).  The solid line shows the fitted signal plus background
shape and the small shaded area the background. The arrows indicate the fit
window. }
\label{fig:mee}
\end{figure}

The electron energy resolution  is described by sampling,
noise, and constant terms. In the Monte Carlo simulation of forward electrons 
we use a sampling term
of $15.7\%/\sqrt{p/\hbox{GeV}}$, derived from beam tests. The noise term is
determined by pedestal distributions derived from the $W$ data sample. We
constrain the constant term to $\cem= 1.0^{+0.6}_{-1.0}\%$ by requiring that
the width of the dielectron invariant mass spectrum predicted by the Monte Carlo
simulation is consistent with the $Z$ data (Fig.~\ref{fig:mee}).
\vskip 1.65 in

Beam tests show that the electron energy response of the calorimeter can be
parameterized by a scale factor $\alphaem$ and an offset $\deltaem$. We
determine these {\it in situ} using 
$\zee$ decays. 
We obtain for forward electrons $\deltaem = -0.1\pm0.7$~GeV and 
$\alphaem =0.95179\pm0.000187$ by fitting the observed mass spectra
while constraining the resonance masses to the measured value of the  $Z$ boson mass ~\cite{ewwg,pdg}. 
The uncertainty in $\alphaem$ is dominated by the finite size of the $Z$
sample. Figure~\ref{fig:mee} shows the observed dielectron mass spectrum from
the dielectron sample, and the line shape predicted by the Monte Carlo simulation for the 
fitted values of $\cem$, $\alphaem$, and $\deltaem$.

\begin{figure}[htb]
\begin{tabular}{ll}
\includegraphics{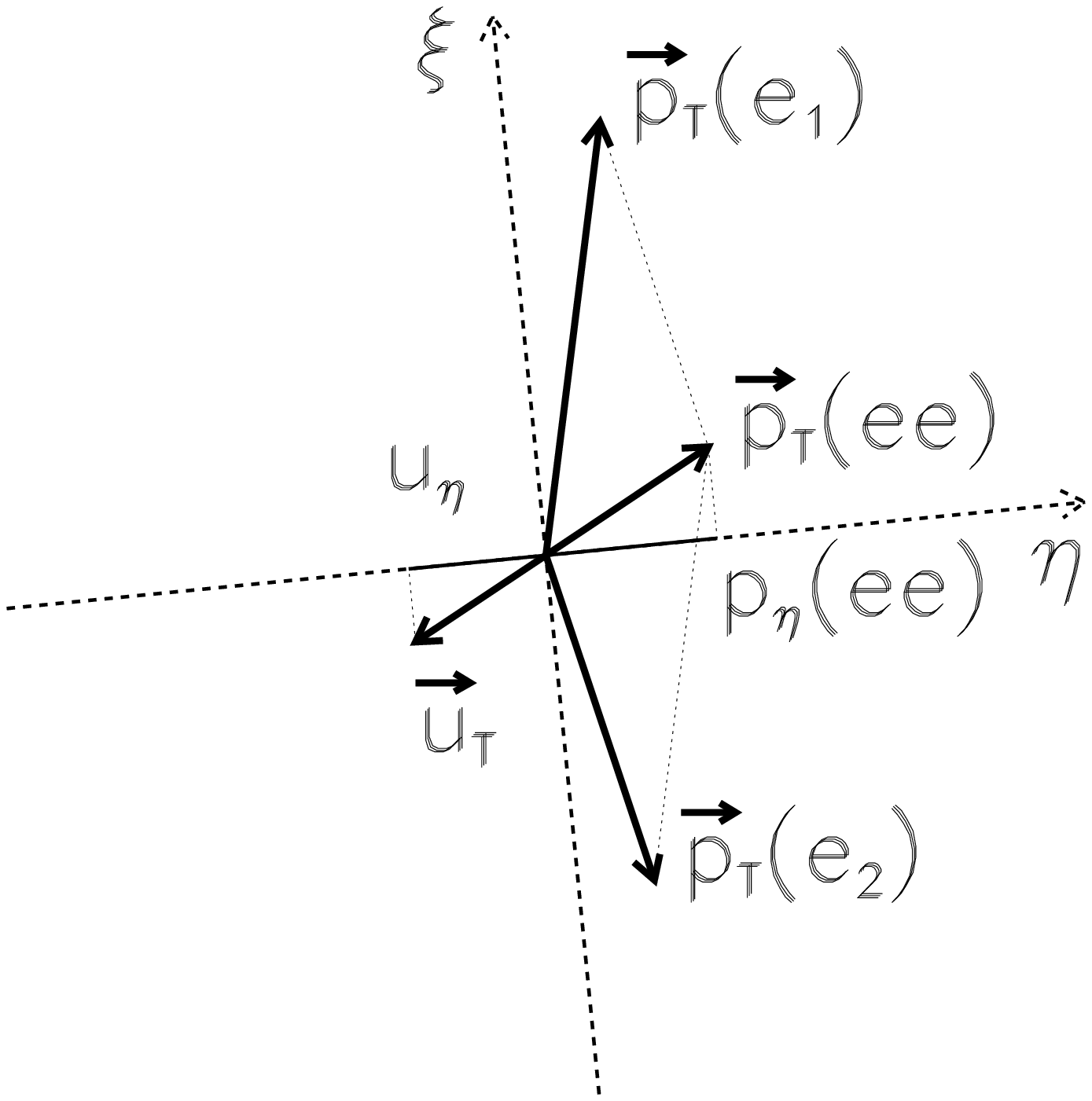}
\includegraphics{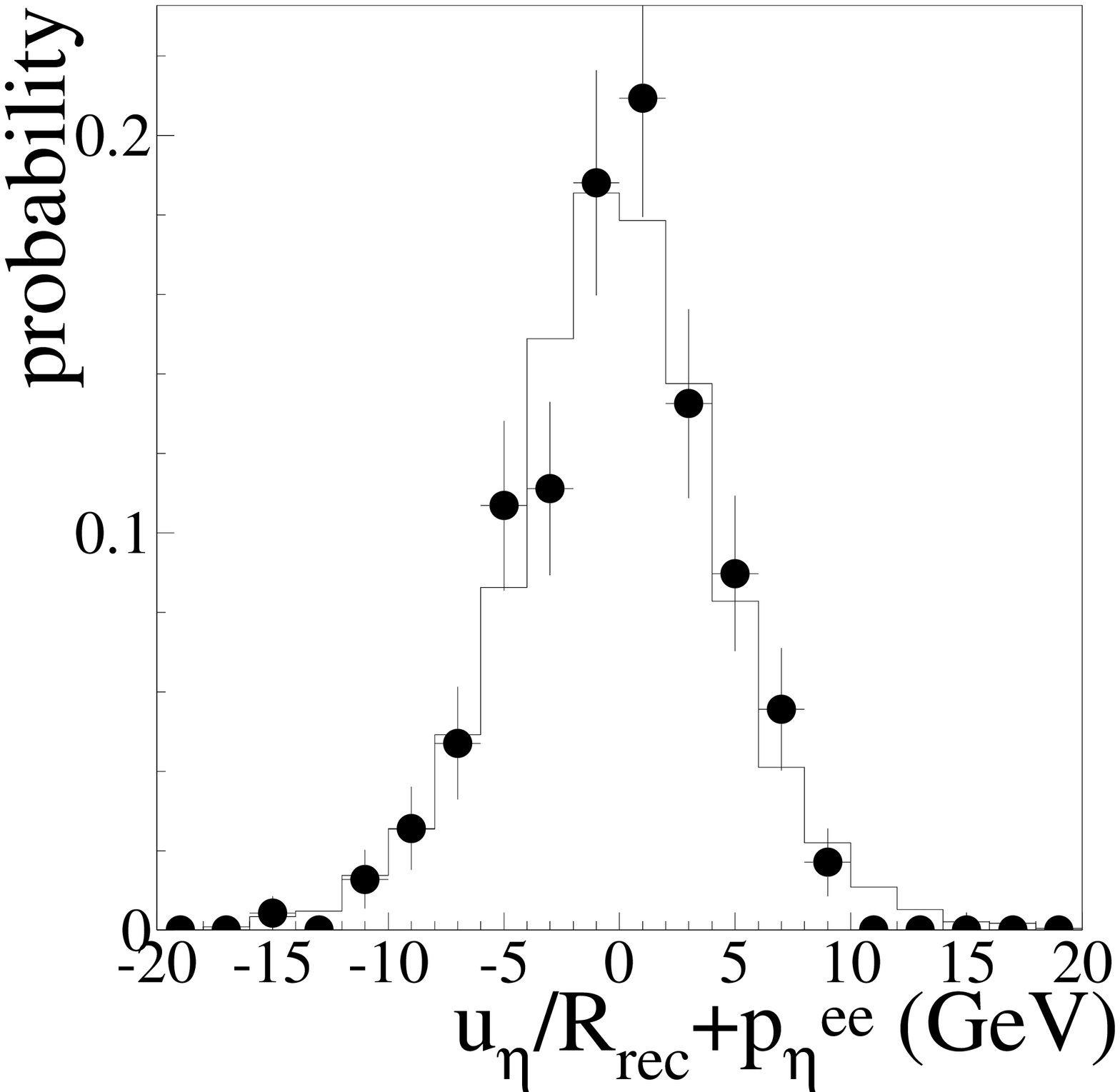}
\end{tabular}
\caption[]{ The definition of the $\eta$-axis (left).  The plot of 
$\ueta/\mbox{R}_{\mbox{\scriptsize rec}}+\pteta$ (right) for the 
data ($\bullet$) and simulation~(---).}
\label{fig:hadres}
\end{figure}

We calibrate the response of the detector to the underlying event, relative to
the EM response, using $Z$ boson data sample.  In $\zee$ decays, momentum conservation requires  
$\pteev=-\utv$, where $\pteev$ is the sum of the two electron $\pt$ vectors.
To minimize sensitivity to the electron energy resolution, we project $\utv$ and
$\pteev$ on the inner bisector of the two electron directions, 
called the $\eta$-axis (Fig.~\ref{fig:hadres}). We call the projections $\ueta$
and $\pteta$.
\vskip 1.6in

Detector simulations based on the \GEAN\ program~\cite{geant} 
predict a detector response to the recoil particle momentum of the form
$\mbox{R}_{\mbox{\scriptsize rec}} = \alpharec + \betarec \ln(\pt/\hbox{GeV})$.
We constrain  $\alpharec$ and $\betarec$ by comparing the mean 
value of $\ueta+\pteta$ with Monte Carlo predictions for different
values of the parameters. We use  $\alpharec=0.693\pm0.060$ and 
$\betarec=0.040\pm0.021$ with a correlation coefficient of~$-0.98$ obtained in central electron analysis ~\cite{wmass1bcc}. We check that Z events with both 
electrons in the end calorimeter give a recoil response measurement  statistically consistent with the above (Figure~\ref{fig:hadres}). 

The recoil momentum resolution has two components. We smear the magnitude of the
recoil momentum with a resolution of $\srec/\sqrt{\pt/\hbox{GeV}}$.
We describe the detector noise and pile-up, which are independent of the boson
$\pt$ and azimuthally symmetric, by adding the $\mpt$
from a random $\ppbar$ interaction, scaled by a factor $\alphamb$, to the
smeared boson $\pt$. To model the luminosity dependence of this resolution
component correctly, the sample of $\ppbar$ interactions was chosen to have the
same luminosity spectrum as the $W$ sample. 
We constrain the parameters by comparing the observed rms of 
$\ueta/\mbox{R}_{\mbox{\scriptsize rec}}+\pteta$ with Monte Carlo predictions.
We use   $\srec=0.49\pm0.14$ and $\alphamb=1.032\pm0.028$ with a
correlation coefficient of $-0.60$ measured in central electron analysis ~\cite{wmass1bcc}. Since we exlude the cells occupied by the electrons, the average transverse energy
flow, $S_T = \sum_i E_i\sin\theta_i$, is  higher for the $W$ sample
than for the $Z$ sample.  
This bias is caused by requiring the identification of 
two electrons in the $Z$ sample versus one in the $W$ sample.
The larger energy flow translates into a 
slightly broader recoil momentum resolution in the $W$ sample. 
$\alphamb$ is corrected by a factor $1.03\pm0.01$ to
account for this effect in the $W$ boson model.
The $p_\eta$ balance
 width is in good agreement between data and Monte Carlo for our Z event sample. 
 Hence we use the same recoil resolution for forward electorn $W$ events as for the central $W$ events
 \cite{wmass1bcc}. 
Figure~\ref{fig:hadres} shows a plot of 
$\ueta/\mbox{R}_{\mbox{\scriptsize rec}}+\pteta$ when both electrons are in the forward calorimeters.

Backgrounds in the $W$ sample are $\wte$ decays (1.0\%, which is included into the Monte Carlo simulation ), 
hadrons misidentified as electrons (3.64\%$\PM$0.78\%, determined from the data ), and $\zee$ decays 
(0.26\%$\PM$0.02\%, determined from HERWIG/GEANT simulations). Their shapes are
included in the probability density functions used in the fits.
\vskip 1.7in
\begin{figure}[htb]
\begin{tabular}{ll}
\includegraphics{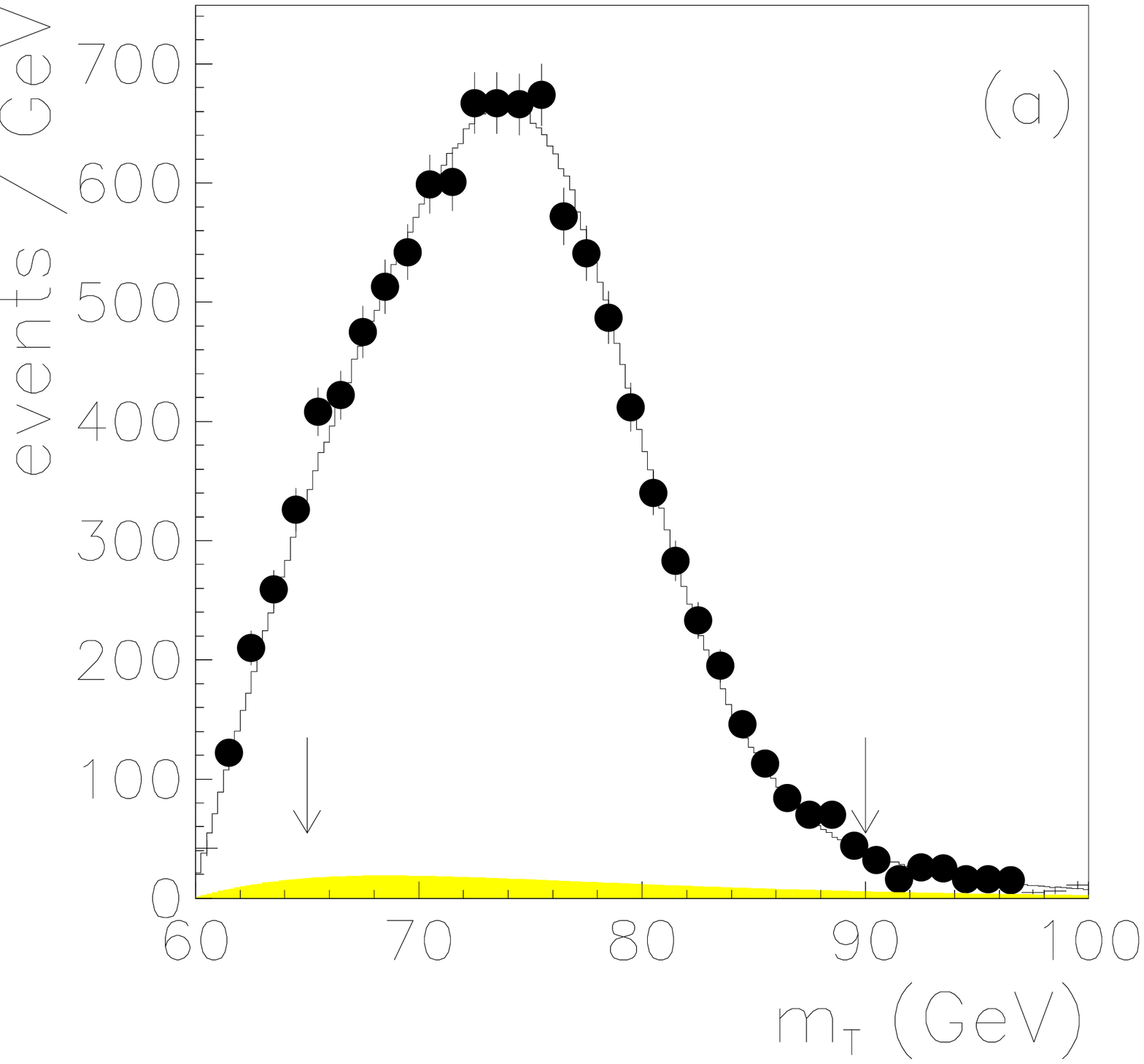}
\includegraphics{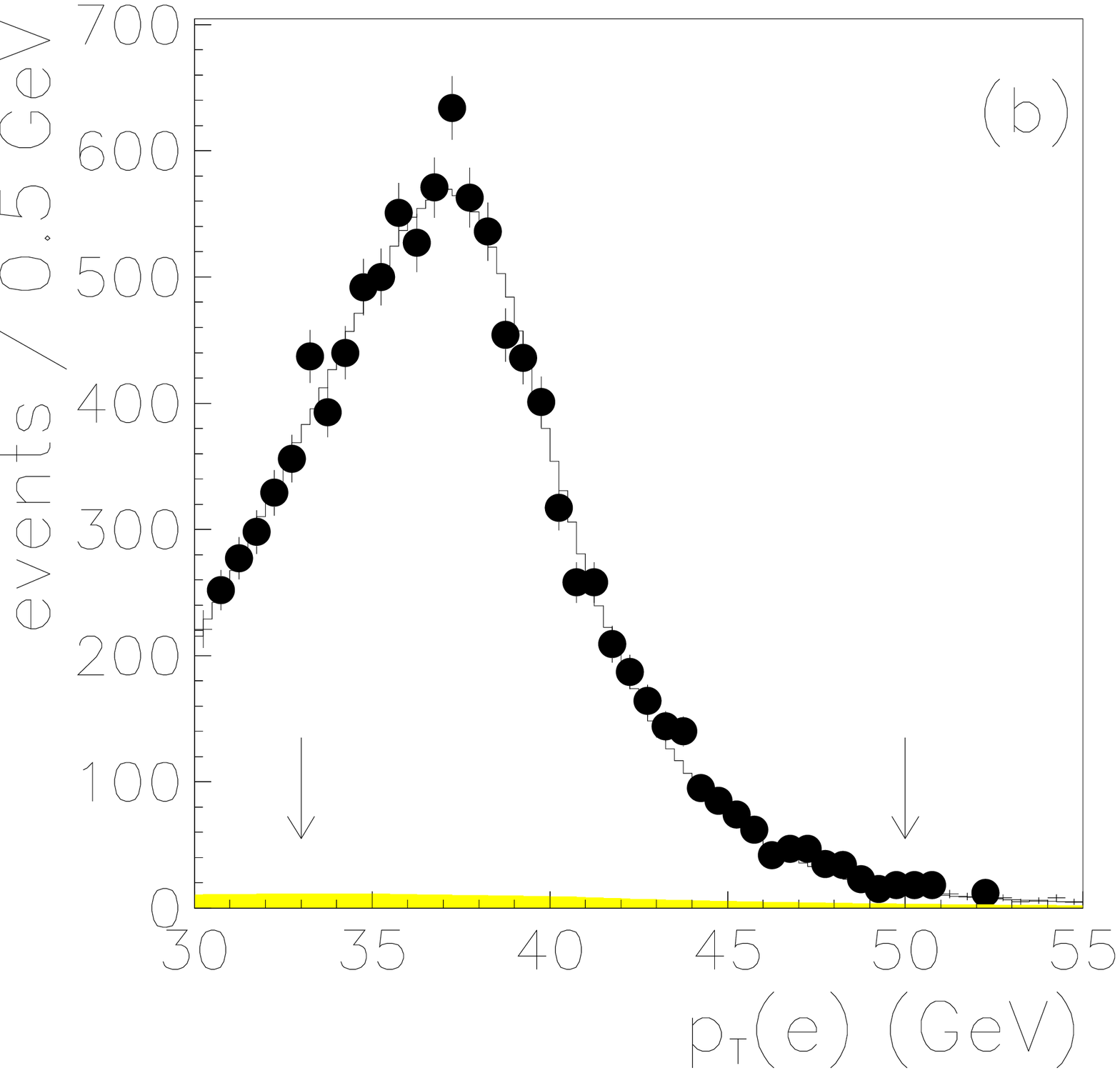}
\includegraphics{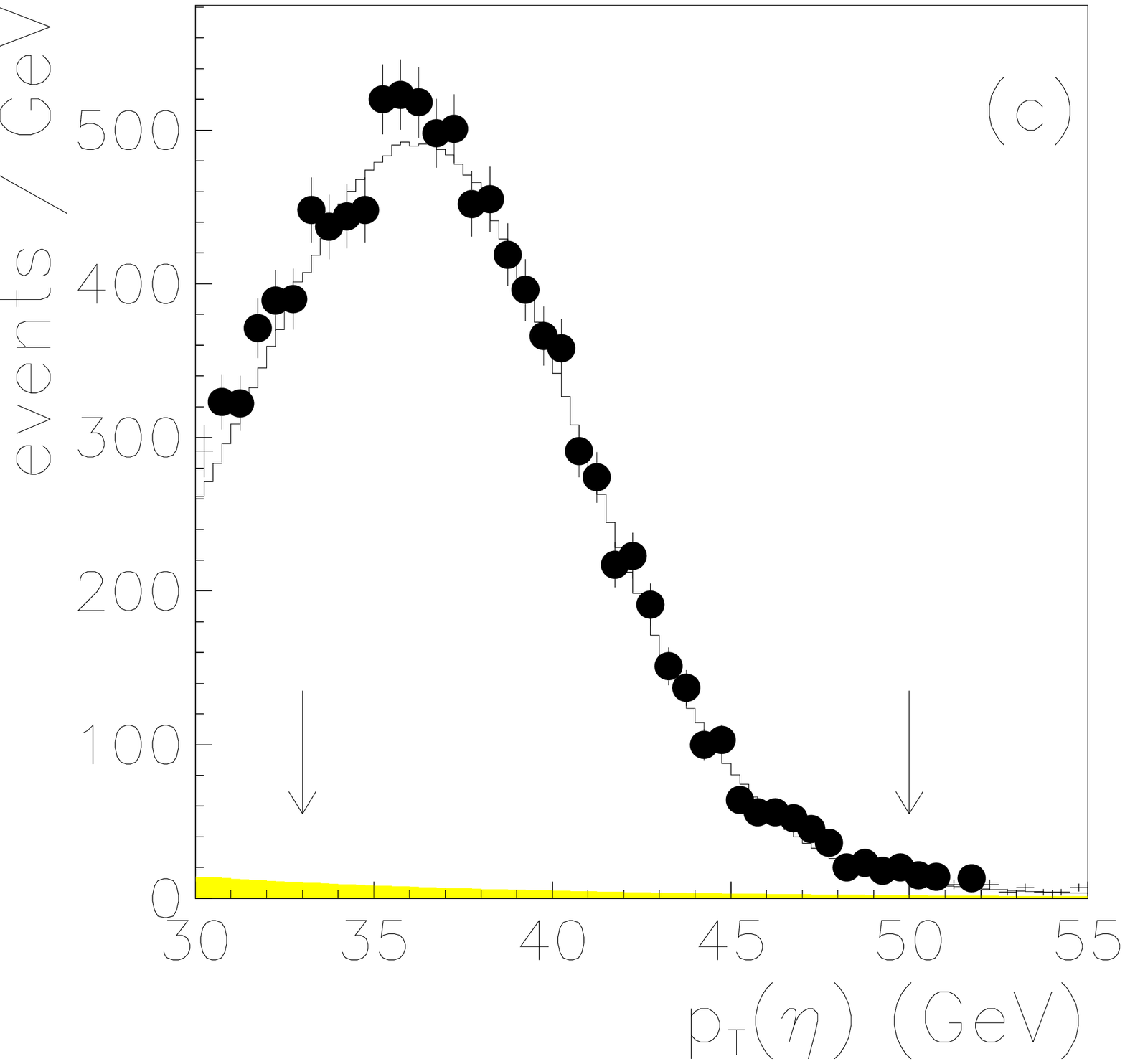}
\end{tabular}
\vskip 0.2cm
\caption[]{ Spectra of (a) $\mt$, (b) $\pte$ and (c) $\ptnu$ from the data ($\bullet$), 
the fit (---), and the backgrounds (shaded). The arrows
indicate the fit windows.}
\label{fig:mt_fit}
\end{figure}

In principle, if the acceptance for the $\wev$ decays were complete, the transverse mass distribution or the lepton $\pt$ distribution would be independent of the $W$ rapidity. However, cuts on the electron angle in the laboratory frame cause the observed distributions of the transverse momenta to depend on the $W$ rapidity. Hence a constraint on the $W$ rapidity distribution is useful in contraining the production model uncertainty on the W mass. We introduce in the Monte Carlo a scale factor $k_\eta$ for the $W$ boson rapidity as $\eta_W \rightarrow k_\eta \eta_W$ and extract it by fitting Monte Carlo electron rapidity distribution simulated with different value of $k_\eta$ to the data. The scale factor is found  to be consistant, within errors, with unity with uncertainty 1.6\%. 
 
The fit to the $\mt$ distribution (Fig.~\ref{fig:mt_fit}(a)) 
yields $\mw=80.766$~GeV with a statistical uncertainty of 108~MeV. A $\chi^2$ test gives $\chi^2=17$ for 25 bins which
corresponds to a confidence level of 81\%.
The fit to the $\pte$ distribution (Fig.~\ref{fig:mt_fit}(b))
yields $\mw=80.587$~GeV with a statistical uncertainty of 125~MeV.
The confidence level of the $\chi^2$ test is 5\%. The fit to the $\ptnu$ distribution (Fig.~\ref{fig:mt_fit}(c)) yields $\mw=80.726$~GeV with a statistical uncertainty of 163~MeV and  confidence level  33\%.

We estimate systematic uncertainties in $\mw$ from the Monte
Carlo parameters by varying them within their uncertainties (Table~I).
In addition to the parameters described above, the calibration 
of the electron polar angle measurement contributes a significant uncertainty.
We use muons from $\ppbar$ collisions and cosmic rays to calibrate the
drift chamber measurements and $\zee$ decays to align the calorimeter 
with the drift chambers. Smaller uncertainties are due to
the removal of the cells occupied by the electron from the computation of 
$\utv$, the uniformity of the calorimeter response, and the 
modeling of trigger and selection biases. 
\vskip -0.5cm
\begin{table}[ht]
\begin{center}
\caption{Uncertainties in the $W$ boson mass measurement in MeV.}
\begin{tabular}{lcc}
\hline
Source                      & Forward & Forward + Central\\
\hline
$W$ boson statistics             & 107 & 61  \\
\hline
$Z$ boson statistics             & 181 & 59  \\
\hline
Calorimeter linearity       &  52 & 25  \\
\hline
Calorimeter uniformity      & --  &  8  \\
\hline
Electron resolution         & 42  & 19  \\
\hline
Electron angle calibration  & 20  & 10 \\
\hline
Recoil response             & 17  & 25  \\
\hline
Recoil resolution           & 42  & 25  \\
\hline
Electron removal            & 2   & 12  \\
\hline
Selection bias              & 5   &  3  \\
\hline
Backgrounds                 & 20  &  9  \\
\hline
Parton distribution functions     & 35  &  15  \\
\hline
Parton luminosity           & 2   &  4  \\
\hline
$\ptw$ spectrum                    & 25  & 15  \\
\hline
$W$ boson width               & 10  & 10  \\
\hline
radiative decays       & 1   & 12  \\
\hline
\end{tabular}
\end{center}
\end{table}

\vskip -0.5cm
The uncertainty due to the model for $W$ boson production and decay consists
of several components (Table~I).
We assign an uncertainty that
characterizes the range of variations in $\mw$ obtained when employing
several recent parton distribution functions: MRST~\cite{mrst}, MRSA~\cite{mrsa}, 
MRSR2~\cite{mrsr2}, CTEQ5M~\cite{cteq5m}, and CTEQ4M~\cite{cteq4m}. We allow 
the $\ptw$ spectrum to vary within constraints derived from the $\ptee$ spectrum
of the $Z$ data and from $\Lambda_{QCD}$~\cite{pdg}. The 
uncertainty due to radiative decays contains an estimate of the effect of
neglecting double photon emission in the Monte Carlo 
simulation~\cite{baur_twophoton}.

The fit to the $\mt$ spectrum results in a $W$ boson mass of 
$80.766\pm0.108(\hbox{stat})\pm0.208(\hbox{syst})\ \hbox{GeV}$, the fit to
the $\pte$ spectrum results in $80.587\pm 0.125(\hbox{stat}) 
\pm0.217(\hbox{syst})\ \hbox{GeV}$, and the fit to
the $\ptnu$ spectrum results in $80.726\pm 0.163(\hbox{stat}) 
\pm0.307(\hbox{syst})\ \hbox{GeV}$. 
The good agreement of the three fits shows that our simulation models 
the $W$ boson production dynamics and the detector response well.
Fits to the data in bins of
luminosity, $\phie$, $\eta(e)$, and $\ut$ do not show evidence for any 
systematic biases.

We  combine all the six measurements ( fits to central electron ~\cite{wmass1bcc} and forward electron $W$ boson events using three techniques ). We obtain the combined 1994-1995 measurement $\mw = 80.487\pm0.096$~GeV. The $\chi^2$ is  4.6/5 dof, with a probability of 46\%. Combining with the 
  measurement from the 1992--93 data gives the 1992--95 data measurement of~
$\mw = 80.474\pm0.093$~GeV. 

\vskip 1.7in
\begin{figure}[htb]
\begin{tabular}{l}
\includegraphics{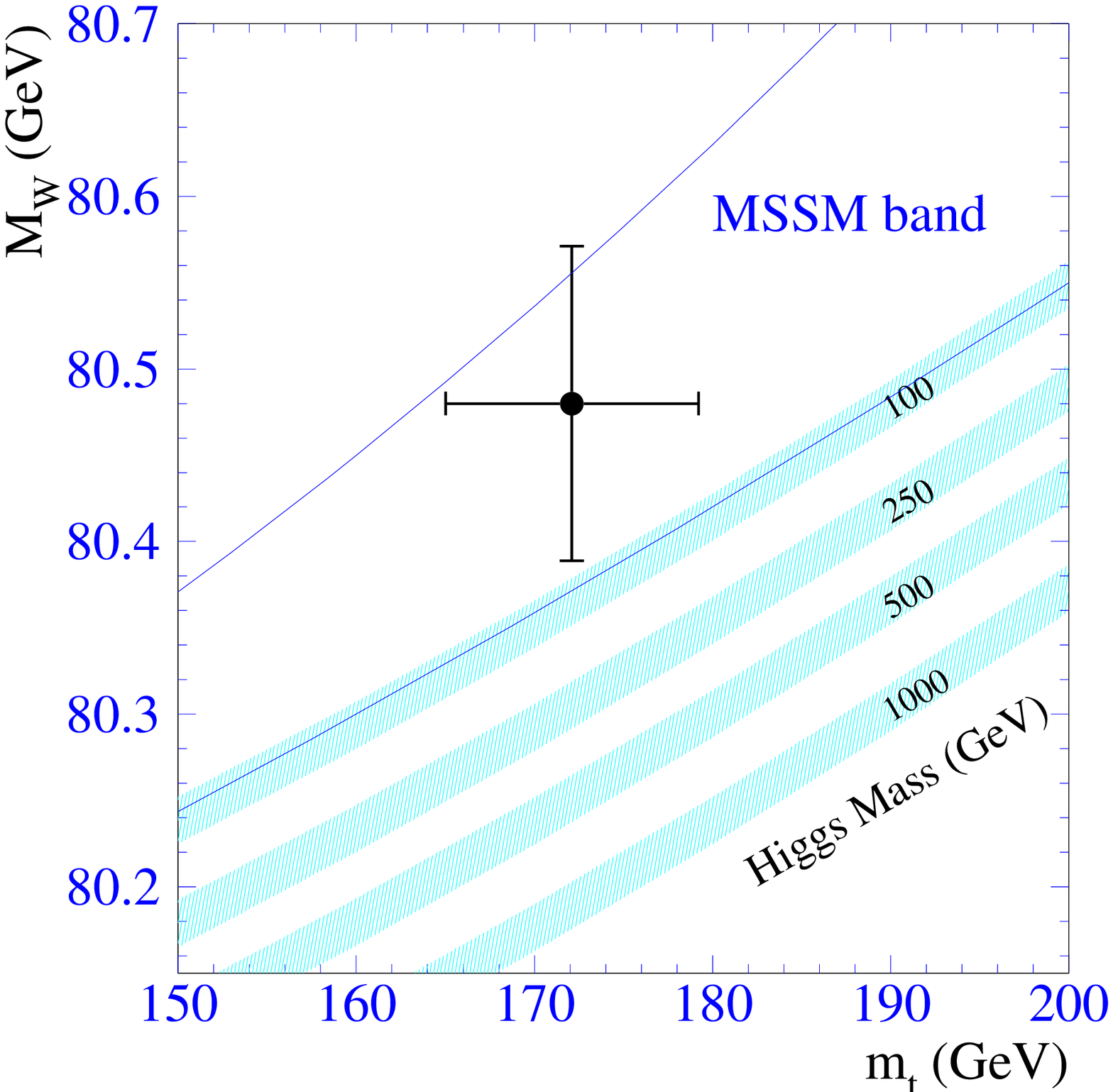}
\end{tabular}
\caption[]{Direct $W$ boson and top quark mass measurements by the 
\Dzero~\protect\cite{d0_top} experiment. The bands show SM
predictions for
the indicated Higgs masses~\protect\cite{mw_v_mt}.} 
\label{fig:higgs}
\end{figure}

Using Eq.~\ref{eq:dr} we find $\Delta r_W  = -0.0317\pm0.0061$, which 
establishes the existence of electroweak corrections to $\mw$ at the 
level of five standard deviations. 
Figure~\ref{fig:higgs} compares the direct measurements of the $W$ boson and
top quark masses to SM predictions. 

%
We thank the Fermilab and collaborating institution staffs for
contributions to this work and acknowledge support from the 
Department of Energy and National Science Foundation (USA),  
Commissariat  \` a L'Energie Atomique (France), 
Ministry for Science and Technology and Ministry for Atomic 
   Energy (Russia),
CAPES and CNPq (Brazil),
Departments of Atomic Energy and Science and Education (India),
Colciencias (Colombia),
CONACyT (Mexico),
Ministry of Education and KOSEF (Korea),
and CONICET and UBACyT (Argentina).


\begin{references}
\bibitem{ep}
    We generically refer to electrons and positrons as electrons.
\end{references}
\end{document}